\documentclass[12pt,a4paper]{article}
\usepackage{epsfig}
\usepackage{a4wide}
\usepackage{amsmath}
\usepackage{bm}
\usepackage{amssymb}
\usepackage{graphicx}
\usepackage{slashed}

\def\xiG{\xi_G}
\def\xiW{\xi_W}
\def\xiB{\xi_B}

\newcommand{\ep}{\ensuremath{\epsilon}} 
\newcommand{\MS}{{\ensuremath{\overline{\mathrm{MS}}}}}

\def\z#1{{\zeta_{#1}}}

\def\T2F{{T^{\,2}_{\! F}}}

\def\Dhat{\ensuremath{\hat{D}}}

\def\qV{\ensuremath{\tilde V}}
\def\qG{\ensuremath{\tilde G}}
\def\qW{\ensuremath{\tilde W}}
\def\qB{\ensuremath{\tilde B}}
\def\bgV{\ensuremath{\hat V}}
\def\bgG{\ensuremath{\hat G}}
\def\bgW{\ensuremath{\hat W}}
\def\bgB{\ensuremath{\hat B}}

\def\ng{n_g}
\def\yu{y_u}
\def\yd{y_d}
\def\yl{y_l}

\def\LG{\ensuremath{\mathcal{L}_{\mathrm{G}}}}
\def\LH{\ensuremath{\mathcal{L}_{\mathrm{H}}}}
\def\LF{\ensuremath{\mathcal{L}_{\mathrm{F}}}}
\def\LFP{\ensuremath{\mathcal{L}_{\mathrm{FP}}}}
\def\LGF{\ensuremath{\mathcal{L}_{\mathrm{GF}}}}

\renewcommand{\text}[1]{#1}

\begin{document}

\boldmath
\title{Gauge coupling beta functions and gauge field anomalous dimensions at four loops in the Standard Model}
\unboldmath

\author{M.~A.~Bezuglov, B.~A.~Kniehl, V.~N.~Velizhanin\\[5mm]
  {\normalsize II. Institut f\"ur Theoretische Physik, Universit\"at Hamburg,}\\
  {\normalsize Luruper Chaussee 149, 22761 Hamburg, Germany}
}
\date{}

\maketitle

\begin{abstract}
We present the beta functions of the gauge couplings and the anomalous dimensions of the gauge fields in the unbroken phase of the Standard Model at four loops.
\end{abstract}

\section{Introduction}
\label{sec:one}

Renormalization group (RG) equations are central tools in quantum field theory, allowing perturbative results obtained at one energy scale to be consistently evolved to another one.
In the Standard Model (SM), they play a crucial role in precision studies of running couplings \cite{Kniehl:2016enc}, threshold matching \cite{Kniehl:2016enc,Sirlin:1985ux,Hempfling:1994ar,Bezrukov:2012sa,Jegerlehner:2012kn,Kniehl:2014yia,Kniehl:2015nwa,Bednyakov:2016onn}, gauge coupling unification scenarios, and the stability of the electroweak vacuum \cite{Bezrukov:2012sa,Degrassi:2012ry,Buttazzo:2013uya,Bednyakov:2015sca}.
Following the discovery of the Higgs boson at the Large Hadron Collider (LHC)~\cite{ATLAS:2012yve,CMS:2012qbp}, the behavior of the SM couplings at very high energies has attracted renewed attention, particularly in connection with vacuum stability analyses.
These studies require increasingly precise RG evolution and, therefore, directly benefit from higher-order knowledge of beta functions.

The computation of RG functions in general renormalizable quantum field theories has a long history.
The two-loop beta functions for all dimensionless couplings were derived by Machacek and Vaughn~\cite{Machacek:1983tz,Machacek:1983fi,Machacek:1984zw} within the $\MS$ renormalization scheme of dimensional regularization.
The complete set of three-loop RG functions of the SM became available in 2012--2013.
Three-loop gauge coupling beta functions were obtained in Refs.~\cite{Mihaila:2012fm,Mihaila:2012pz,Bednyakov:2012rb}, followed by Yukawa coupling beta functions~\cite{Chetyrkin:2012rz,Bednyakov:2012en} and Higgs sector RG functions~\cite{Chetyrkin:2013wya,Bednyakov:2013eba}.
At this order, nontrivial contributions involving the chirality matrix $\gamma_5$ first appear in the Yukawa sector, requiring a careful treatment within dimensional regularization~\cite{Jegerlehner:2000dz}.

The first four-loop SM results were obtained for the strong-coupling beta function~\cite{Bednyakov:2015ooa,Zoller:2015tha}.
At this order, the well-known ambiguity associated with $\gamma_5$ in dimensional regularization affects the gauge sector itself.
A detailed analysis performed in Ref.~\cite{Bednyakov:2015ooa} identified a consistent prescription, which was later independently confirmed in Ref.~\cite{Poole:2019txl} by using Weyl consistency conditions (WCCs) \cite{Osborn:1991gm,Antipin:2013sga,Jack:2013sha,Poole:2019kcm}.
WCCs relate gauge, Yukawa, and scalar RG functions at different loop orders.
In particular, ambiguous four-loop gauge contributions were connected to unambiguous three-loop Yukawa beta functions, providing a powerful consistency check of the four-loop results.
These developments subsequently led to the determination of the complete four-loop gauge sector beta functions from WCC-based analyses~\cite{Davies:2019onf,Bednyakov:2021qxa} and stimulated the development of general high-order methods based on tensor structure decompositions and WCCs \cite{Steudtner:2020tzo,Steudtner:2021fzs,Davies:2021mnc,Steudtner:2024teg}.

In this paper, we extend the direct three-loop calculation of the SM gauge coupling beta functions in Ref.~\cite{Bednyakov:2012rb} to four loops.
We present the four-loop anomalous dimensions of the SM gauge fields and derive the corresponding four-loop gauge coupling beta functions.
The calculation is performed in the background field gauge~\cite{Abbott:1980hw,Abbott:1981ke}, where Ward identities relate the renormalization of background gauge fields directly to the renormalization of the gauge couplings.
As for the treatment of chiral traces, we employ the reading point prescription for $\gamma_5$~\cite{Korner:1991sx}, following Ref.~\cite{Bednyakov:2015ooa}.
We find that the final results are independent of the auxiliary choices introduced by this procedure.

Our calculation provides the first direct four-loop determination of the complete gauge sector renormalization of the unbroken Standard Model and independently confirms the results previously obtained using WCCs \cite{Davies:2019onf,Bednyakov:2021qxa}.
The resulting beta functions constitute an essential ingredient for future precision studies of SM running couplings and vacuum stability at very high energy scales.

This paper is organized as follows.
In Section~\ref{sec:two}, we set up our theoretical framework.
In Section~\ref{sec:three}, we give details of our calculations.
In Section~\ref{sec:four}, we present our four-loop results for gauge coupling beta functions and gauge field anomalous dimensions in Landau gauge.
Section~\ref{sec:five} contains our conclusions.
In an ancillary file published along with this article, we present, in computer-readable form, all renormalization constants and anomalous dimensions in $R_\xi$ gauge.

\section{Background fields and renormalization}
\label{sec:two}

We begin by summarizing the SM Lagrangian formulated within the background field method, closely following the presentation in Ref.~\cite{Denner:1994xt}.
However, in contrast to  Ref.~\cite{Denner:1994xt}, background fields are introduced here exclusively for the gauge sector.
Furthermore, all dimensionful parameters, including masses, are omitted throughout this study, as already stated in Section~\ref{sec:one}.

The Lagrangian employed in our analysis can be written as
\begin{equation}
{\mathcal L} =
  \LG 
+ \LH 
+ \LF 
+ \LGF 
+ \LFP,
\end{equation}
where the individual contributions correspond to the gauge, scalar, fermion, gauge fixing, and ghost sectors, respectively.

The gauge field contribution is given by the Yang--Mills Lagrangian,
\begin{eqnarray}
	\LG & = &
		-\frac{1}{4} G^a_{\mu\nu} G^a_{\mu\nu}
	        -\frac{1}{4} W^i_{\mu\nu} W^i_{\mu\nu}
	        -\frac{1}{4} B_{\mu\nu} B_{\mu\nu},
\end{eqnarray}
with the field strength tensors defined as
\begin{eqnarray}
	G^a_{\mu\nu} & = & \partial_\mu G^a_\nu - \partial_\nu G^a_\mu + g_s f^{abc} G_\mu^b G_\nu^c, \\
	W^i_{\mu\nu} & = & \partial_\mu W^i_\nu - \partial_\nu W^i_\mu + g_2 \epsilon^{ijk} W^j_\mu W^k_\nu, \\
	B_{\mu\nu} & = & \partial_\mu B_\nu - \partial_\nu B_\mu.
\end{eqnarray}
The gauge fields are decomposed into quantum and background components according to
$G^a_\mu = \qG^a_\mu + \bgG^a_\mu$, $W^i_\mu = \qW^i_\mu + \bgW^i_\mu $, and $B_\mu = \qB_\mu + \bgB_\mu$, where the indices $a=1,\ldots,8$ and $i=1,2,3$ label the generators of SU(3) and SU(2), respectively.
We collectively denote the quantum gauge fields by $\qV=\qG,\qW,\qB$ and their background counterparts by $\bgV=\bgG,\bgW,\bgB$.
The associated gauge couplings are $g_s$, $g_2$, and $g_1$. 
The group structure constants enter through the commutation relations,
\begin{equation}
	\left[ T^a, T^b \right]  =  i f^{abc} T^c,\qquad \left[ \tau^i, \tau^j \right]  =  i \epsilon^{ijk} \tau^k,
\end{equation}
	with $T^a = \lambda^a/2$ and $\tau^i = \sigma^i/2$ being color and weak-isospin generators.

The covariant derivative acting on a field that is charged under all the gauge groups looks like
\begin{equation}
	D_\mu = \partial_\mu - i g_s T^a G^a_\mu - i g_2 \tau^i W^i_\mu + i g_1 \frac{Y_W}{2} B_\mu.
\end{equation}
Whenever a field is neutral with respect to one of the gauge factors, the corresponding term is absent.
Using this derivative, the Higgs and fermion sectors can be expressed as
\begin{eqnarray}
	\LH & = & \left( D_\mu \Phi \right)^\dagger  \left( D_\mu \Phi \right)
	- \lambda \left( \Phi^\dagger \Phi \right)^2,\\
	\LF & = &
	\sum\limits_{i=1,2,3}
	    \bigg( i \bar{Q}^L_i  \Dhat Q^L_i  + i \bar L^L_i  \Dhat  L^L_i 	  
+  i \bar{u}^R_g  \Dhat  u^R_g +  i \bar{d}^R_g  \Dhat  d^R_g +  i \bar l^R_g  \hat D  l^R_g  \bigg)\nonumber\\
	& - & \sum\limits_{i,j=1,2,3} \bigg(
	Y^{ij}_u (Q^L_i \Phi^c) u^R_j
	+ Y^{ij}_d (Q^L_i \Phi) d^R_j  + Y^{ij}_l (L^L_i \Phi) l^R_j
	+ \mathrm{h.c.} \bigg).
\end{eqnarray}
Here, $i$ and $j$ enumerate the three fermion generations.
The parameter $\lambda$ denotes the Higgs self-coupling, while $Y_u$, $Y_d$, and $Y_l$ represent the Yukawa matrices.
The left-handed quark $Q_g^L = (u_g, d_g)^L$ and lepton fields $L_g^L=(\nu_g,l_g)^L$ transform as SU(2) doublets, whereas the right-handed quarks ($u^R_g, d^R_g$)  and charged leptons $l^R_g$ are SU(2) singlets.
The Higgs doublet $\Phi$ with $Y_W = 1$ has the following decomposition in terms of the component fields:
\begin{equation}
	\Phi =
	\left(
	\begin{array}{c}
		\phi^+ \\ \frac{1}{\sqrt 2} \left( h + i \chi \right)
		\end{array}
	\right),
	\qquad
	\Phi^c = i\sigma^2 \Phi^\dagger =
	\left(
	\begin{array}{c}
		\frac{1}{\sqrt 2} \left( h - i \chi \right) \\
		-\phi^-
		\end{array}
	\right),
\end{equation}
where $\Phi^c$ denotes the charge-conjugated Higgs doublet with hypercharge $Y_W=-1$.

The gauge fixing sector is implemented exclusively for the quantum gauge fields and is described by
\begin{equation}
	\LGF  =
	-\frac{1}{2\xiG} G_G^a G_G^a
	-\frac{1}{2\xiW} G_W^i G_W^i
	-\frac{1}{2\xiB} G_B^2,
\end{equation}
with
\begin{eqnarray}
	G^a_G  & = & \partial_\mu \qG^a_\mu + g_s f^{abc} \bgG^b_\mu \qG^c_\mu\,, \nonumber\\ 
	G^i_W  & = & \partial_\mu \qW^i_\mu + g_2 \epsilon^{ijk} \bgW^j_\mu \qW^k_\mu\,, \nonumber\\ 
	G_B & = &  \partial_\mu \qB_\mu \, .  
\label{eq:gauge_fix_func}
\end{eqnarray}
In comparison with the conventional linear gauge, the ordinary derivatives are promoted to background field covariant derivatives.
Consequently, the gauge fixing procedure preserves the invariance of the effective action under background gauge transformations.

The Fadeev--Popov part of the Lagrangian is given by
\begin{equation}
	\LFP  =  - \bar c_\alpha \frac{\delta G_\alpha}{\delta \theta^\beta} c_\beta,
\end{equation}
where $\alpha,\beta = G,W,B$, and $\delta G_\alpha/\delta \theta^\beta$ is the variation of the gauge fixing functions in Eq.~\eqref{eq:gauge_fix_func}
under the following infinitesimal quantum gauge transformations:
\begin{eqnarray}
	\delta \qG^a_\mu & = & (D_\mu \theta_G)^a = \partial_\mu \theta_G^a + g_s f^{abc} G^b_\mu \theta_G^c,\nonumber\\
	\delta \qW^i_\mu & = & (D_\mu \theta_W)^i = \partial_\mu \theta_W^i + g_2 \epsilon^{ijk} W^j_\mu \theta_W^k,\nonumber\\
	\delta \qB_\mu & = & \partial_\mu \theta_B.
	\label{eq:gt}
\end{eqnarray}
It should be noted that the covariant derivatives appearing in Eq.~\eqref{eq:gt} are constructed from the complete gauge fields $V = \qV + \bgV$.
The corresponding background gauge transformations are obtained straightforwardly by replacing the full gauge fields with their background components only, i.e. $V\to\bgV$.

The background field formulation preserves the gauge invariance of the effective action with respect to transformations of the background fields.
As a consequence, one can derive Ward identities analogous to those familiar from QED.
These identities imply a particularly simple connection between the renormalization constants of the gauge couplings and those of the background gauge fields,
\begin{equation}
	Z_{g_i} = Z_{\bgV_i}^{-1/2}, \qquad i=1,2,3,
	\label{eq:bftog}
\end{equation}
where $Z_{\bgV_i}$ denotes the renormalization constant associated with the background gauge fields $\bgV^\mu_i = \bgB^\mu, \bgW^\mu, \bgG^\mu$, while $Z_{g_i}$ corresponds to the renormalization of the SM gauge couplings $g_i = g_1,g_2, g_s$.

Since the calculation is performed retaining the complete dependence on the gauge fixing parameters $\xi_i = \xiB, \xiW, \xiG $, their renormalization must also be determined.
Owing once again to the Ward identities, radiative corrections do not modify the longitudinal components of the propagators for the quantum gauge fields.
This property directly leads to the relation
\begin{equation}
	Z_{\xi_i} = Z_{\qV_i},	
	\label{eq:ztoxi}
\end{equation}
where $Z_{\xi_i}$ represents the renormalization constants of the gauge fixing parameters.
The quantum gauge fields $\qV_i$ are renormalized within the \MS\ scheme through the field renormalization constants $Z_{\qV_i}$.

Equations~(\ref{eq:bftog}) and (\ref{eq:ztoxi}) show that the determination of the renormalization constants requires the computation of gauge boson self-energies involving both the quantum fields $\qV$ and the corresponding background fields $\bgV$.

For the calculation of the renormalization constants, following Ref.~\cite{Larin:1993tp} (see also Refs.~\cite{Tarasov:1976ef,Vladimirov:1979zm,Tarasov:1980au}), we exploit the multiplicative renormalizability of the corresponding Green's functions.
 The renormalization constants $Z_{V}$ relate the dimensionally
 regularized one-particle-irreducible two-point functions $\Gamma_{V,\mathrm{Bare}}$ with the renormalized ones $\Gamma_{V,\mathrm{Ren}}$ as
\begin{equation}
   \Gamma_{V,\mathrm{Ren}}\left(\frac{Q^2}{\mu^2},a_i,\xi_k\right)=\lim_{\ep \rightarrow 0}
   Z_{V} \left(\frac{1}{\ep},a_i,\xi_k\right)
   \Gamma_{V,\mathrm{Bare}}\left(Q^2,a_{i,\mathrm{Bare}},\xi_{k,\mathrm{Bare}},\ep\right),
\label{eq:Gamma}   
\end{equation}
  where $a_{i,\mathrm{Bare}}$ and $\xi_{k,\mathrm{Bare}}$ denote the bare parameters of the theory.
  For convenience, we introduce the following notation
 \begin{eqnarray}
	 a_i  & = &  \left( \frac{g_1^2}{16\pi^2}, \frac{g_2^2}{16\pi^2}, \frac{g_s^2}{16\pi^2}, \frac{Y_u^2}{16\pi^2}, \frac{Y_d^2}{16\pi^2}, \frac{Y_l^2}{16\pi^2}, \frac{\lambda}{16\pi^2} \right),
 \end{eqnarray}
	and $\xi_{k}=\xiG, \xiW, \xiB$.

As for the gauge couplings, the bare parameters are related to the renormalized ones in the \MS\ scheme by the following formula:
\begin{equation}
	a_{k,\mathrm{Bare}}\mu^{-\epsilon} = Z_{a_k} a_k(\mu)=a_k+\sum_{n=1}^\infty c_k^{(n)}\frac{1}{\epsilon^n}.
	\label{eq:bare}
\end{equation}
In order to extract the four-loop contribution to $Z_V$ from the corresponding self-energies, it is sufficient to know the four-loop renormalization constants for the gauge couplings, the three-loop results for the Yukawa couplings, and the two-loop result for the scalar quartic coupling $\lambda$.
This is due to the fact that the Yukawa vertices appear for the first time in the two-loop self-energies and the Higgs self-coupling enters the result only at the third order of perturbation theory.

The four-dimensional beta functions, denoted by $\beta_i$, are defined via
\begin{equation}
		\beta_i(a_k) = \mu^2\frac{d a_i(\mu,\epsilon)}{d \mu^2}\bigg|_{\epsilon=0} .		
		\label{eq:beta}
\end{equation}
Given the fact that the bare parameters do not depend on the renormalization scale $\mu$, the expressions for $\beta_i$ can be obtained~\cite{Machacek:1983tz} by differentiation of Eq.~\eqref{eq:bare} with respect to $\ln \mu^2$ and taking into account only the leading order of the expansion in $\epsilon$.
We so obtain
	\begin{equation}
 	\beta_k =  \sum_{l}a_l\frac{\partial c_k^{(1)}}{\partial a_l}-c_k^{(1)}.
	\end{equation}

In \MS-like schemes, the renormalization constants for the Green's functions may be expanded as
\begin{equation}
  Z_{\Gamma} = 1 + \sum\limits_{k=1}^{\infty} \frac{Z_{\Gamma}^{(k)}}{\ep^k}.
\label{eq:ren_const_ep_decom} 
\end{equation}
Differentiating Eq.~\eqref{eq:ren_const_ep_decom} with respect to $\ln \mu^2$, we obtain the following all-order expression for the anomalous dimensions:
\begin{equation}
	\gamma_\Gamma\equiv - \mu^2 \frac{\partial \ln Z_\Gamma}{\partial \mu^2}
	=  - \left[\sum\limits_{j} \big(\beta_j - a_j\epsilon\big)\frac{\partial Z_\Gamma}{\partial a_j}\right] Z_\Gamma^{-1}.
\end{equation}
It turns out that the above expression is finite as $\ep\to0$, so that 
\begin{equation}
	\gamma_\Gamma= \sum\limits_j a_j \frac{\partial Z^{(1)}_\Gamma}{\partial a_j}.
\end{equation}

Both the gauge coupling beta functions $\beta_i$ and the gauge field anomalous dimensions $\gamma_i$ can be expanded perturbatively, as
\begin{eqnarray}
\beta_i=-\sum_n a_i\, \beta_i^{(n)}(a_k) ,
\qquad
\gamma_i=-\sum_n a_i\, \gamma_i^{(n)}(a_k,\xi_k),
\end{eqnarray}
where $n$ is the order of perturbative theory, i.e., the number of loops.
In the following, we will compute the gauge coupling beta functions $\beta_i^{(4)}$ and the gauge field anomalous dimensions $\gamma_i^{(4)}$ at four loops.

\section{Details of calculations}
\label{sec:three}

In order to calculate the bare two-point functions of the quantum and background fields, we use \texttt{DIANA} \cite{Tentyukov:1999is}, which generates all the Feynman diagrams with the help of \texttt{qgraf}~\cite{Nogueira:1991ex} with Feynman rules from Ref.~\cite{Bednyakov:2013eba} and calls \texttt{FORM} \cite{Vermaseren:2000nd} to compute the obtained expressions diagram by diagram.
The color algebra is manipulated using \texttt{COLOR} \cite{vanRitbergen:1998pn}, and the Feynman integrals are calculated with \texttt{FORCER}~\cite{Ruijl:2017cxj}.
All computations are performed in arbitrary gauge, i.e., we keep gauge fixing parameters for all three gauge fields.
The numbers of calculated diagrams are given in Table~\ref{tab:Dia}.

\renewcommand{\arraystretch}{1.5}

\begin{table}
\begin{center}
\begin{tabular}{| c | c | c | c | c | c | c | c |}
\hline
         &  1 &   2 &     3 & 4         &   $g^8$   & Mno1     & Mhaha     \\ \hline
$\bgW_i$ & 11 & 389 & 36647 & 2903602   & 9900      & 1488     & 744       \\ \hline
$\bgB$   &  6 & 214 & 20144 & 2813358   & 9900      & 1488     & 744       \\ \hline
$\bgG$   &  4 &  73 &  4183 & 438211    & 5004      & 216      & 108       \\ \hline
\end{tabular}
\caption{Numbers of self-energy diagrams with external gauge fields in the unbroken SM at one, two, three, and four loops are listed in the first four columns; the fifth column gives the numbers of four-loop diagrams with eight gauge vertices (order $g^8$) as in Fig.~\ref{fig:dia}(b); the sixth and seventh columns show the numbers of four-loop diagrams with four gauge verices and four Yukawa vertices as in Fig.~\ref{fig:dia}(a) for the different types of \texttt{FORCER} topologies that yield non-zero contributions.
}
\label{tab:Dia}
\end{center}
\end{table}

In the first step, we use the ``na{\"\i}ve'' prescription for $\gamma_5$, that is, we set all fermion traces containing an odd number of $\gamma_5$ matrices to zero.
However, starting at four loops, traces with an odd number of $\gamma_5$ matrices can contribute.
It was observed that, even in the gaugeless limit, this additional contribution depends on the prescription used for $\gamma_5$.
In Ref.~\cite{Bednyakov:2015ooa}, it was shown that the reading point method can yield a reasonable result if the reading point is chosen to be one of the internal vertices.
The correctness of this treatment was confirmed by the application of WCCs, which allow one to relate the renormalization of the problematic four-loop gauge beta functions to the three-loop Yukawa beta functions.
Therefore, to complete our calculation, we need to re-evaluate, in a second step, all diagrams with an odd number of $\gamma_5$ matrices in two fermion traces.

After the first step, using the ``na{\"\i}ve'' $\gamma_5$ prescription, which does not take into account terms with an odd number of $\gamma_5$ matrices in fermion traces, we obtain the following differences between our incomplete expressions for the four-loop gauge beta functions $\hat\beta_i$ and the results of Refs.~\cite{Davies:2019onf,Bednyakov:2021qxa}:
\begin{eqnarray}
\hat{\beta}_S-\beta_S&=&-4 a_3^2 (\yu - \yd)^2 (1 + 6 \z3),\label{eq:diffGS}\\
\hat{\beta}_2-\beta_2&=&
-\frac{1}{6} a_2 \bigg\{
a_2 \Big(3 \yu -3 \yd - \yl\Big)^2 (10 - 3 \z3) 
+ a_1 \Big[
9 \yu^2 (2 + 3 \z3) 
+ 3 \yd^2 (4 - 3 \z3) \nonumber\\
&&- 2 \yd (3 \yu - \yl) (4 - 3 \z3) 
+ \yl^2 (4 + 15 \z3) 
- 2 \yu \yl (8 + 21 \z3)
\Big]
\bigg\},\label{eq:diffG2}
\\
\hat{\beta}_1-\beta_1&=&
-\frac{1}{18} a_1 \bigg\{
a_1 \Big[
 \yu^2 (134 + 561 \z3) 
 + \yd^2 (38 - 15 \z3) 
 + 3 \yl^2 (34 + 195 \z3) 
\nonumber\\
&&
 - 2 \yu\yd  (50 + 57 \z3) 
 - 2 \yu\yl  (110 + 579 \z3) 
 + 2 \yd\yl (26 + 75 \z3)
 \Big]
\nonumber\\
&&
+ 9 a_2 \Big[
  9 \yu^2 (2 + 3 \z3) 
 + 3 \yd^2 (4 - 3 \z3) 
 + \yl^2 (4 + 15 \z3) 
 - 2  \yu \yl(8 + 21 \z3)
\nonumber\\
&&
 -  2 \yd (3 \yu-\yl) (4 - 3 \z3) 
 \Big]
 \bigg\}.\label{eq:diffG1}
\end{eqnarray}

In the second step, using \texttt{DIANA} \cite{Tentyukov:1999is}, we extract all diagrams with at least two fermion traces, excluding diagrams in which the fermion traces appear in vector or scalar self-energy insertions.
The total numbers of selected diagrams for the different vector fields are given in Table~\ref{tab:Dia}.

According to our Feynman rules, $\gamma_5$ appears in all triple vertices containing fermion lines, including the QCD vertex, for which we use $(1+\gamma_5)/2+(1-\gamma_5)/2$.
This means that all such vertices in the diagram contain $\gamma_5$.
When applying the reading point method, we isolate $\gamma_5$ from the chosen reading point vertex, while, for the remaining appearances of $\gamma_5$ inside the considered fermion trace, we use the rules
\begin{equation}
\{\gamma_5,\gamma_{\mu}\}=0,\qquad\gamma_5^2=1,
\end{equation}
to eliminate an even number of $\gamma_5$ matrices.

\begin{figure}
\begin{center}
\begin{tabular}{ccc}
\multicolumn{3}{c}{\includegraphics[width=1.00\textwidth]{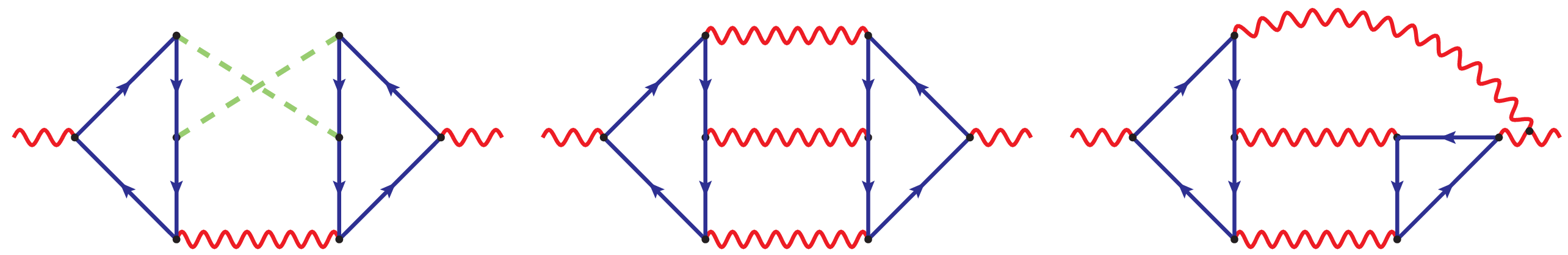}}\\
\hspace {24mm}(a) &\hspace {43.5mm} (b) &\hspace {20mm} (c) 
\end{tabular}
\caption{Typical Feynman diagrams with odd numbers of $\gamma_5$ matrices inside fermion traces yielding non-zero contributions.
Diagram~(a) contributes at total gauge coupling order 4 and Yukawa coupling order 4; diagram~(b) contributes at total gauge coupling order 8; and diagram~(c) contributes at total gauge coupling order 6 or higher.
  Wavy lines denote vector fields $(G,W,B)$, dashed lines denote scalar fields $(\varphi^+,\varphi^{-},h,\chi)$, and solid lines with arrows denote fermion fields $(u,d,l)$.}
\label{fig:dia}
\end{center}
\end{figure}

We evaluate all selected diagrams using the $\gamma_5$ prescription described above and find that only the diagrams with \texttt{FORCER} \cite{Ruijl:2017cxj} topologies \texttt{Mno1} and \texttt{Mhaha} yield non-zero contributions;
examples of such diagrams are shown in Figs.~\ref{fig:dia}(a) and (b).
In principle, there are a lot of non-zero diagrams which contain fermion triangles with three vector lines attached to them (see, e.g., Fig.~\ref{fig:dia}(c)), but, as is well known, such contributions add up to zero when we take into account all SM fields and put the number of colors to be $N_c=3$.

As in previous analogous calculations~\cite{Bednyakov:2015ooa}, we obtain different results depending on the choice of reading point for diagrams such as those shown in Figs.~\ref{fig:dia}(a) and (b).
However, if internal vertices are chosen as the reading points and the results are summed over all possible choices and divided by the number of internal vertices (in our case, three internal vertices for each fermion trace, giving an overall factor of $3\times 3=9$), we exactly recover the excess contributions in Eqs.~(\ref{eq:diffGS})--(\ref{eq:diffG1}).

In Ref.~\cite{Bednyakov:2015ooa}, it was argued that choosing internal vertices as reading points is preferable, since this renders the finite parts of the final results gauge independent.
This choice is further supported by the calculation using tadpoles, where the treatment of the $\gamma$ algebra and the integration can be separated.
Moreover, the corresponding four-loop contributions satisfy the WCCs~\cite{Osborn:1991gm,Antipin:2013sga,Jack:2013sha,Poole:2019kcm}, providing an independent confirmation of the reading point prescription adopted in Ref.~\cite{Bednyakov:2015ooa}.

Notice that the total contribution from all diagrams of the type shown in Fig.~\ref{fig:dia}(b) vanishes; in other words, if no scalar fields are present in such $\gamma_5$-problematic diagrams, their sum is zero.

\section{Results}
\label{sec:four}

Our final results for the four-loop coefficients of the SM gauge coupling beta functions read:
\allowdisplaybreaks
\begin{eqnarray}
\beta_3^{(4)}&&=
a_3^4 \bigg(
+\frac{149753}{6}
+3564 \z3
-\ng \bigg(\frac{1078361}{81}
+\frac{13016}{27}\z3\bigg)
+\ng^2 \bigg(
\frac{100130}{81}
+\frac{25888}{81}\z3\bigg)
\nonumber\\&&
+\frac{8744}{729}\ng^3
\bigg)
+a_3^3 (\yu+\yd) \bigg(\frac{9959}{9}-272 \z3- \frac{3250}{27}\ng\bigg)
-a_3^3a_2 \ng \bigg(\frac{713}{36}+\frac{265}{3}\z3
\nonumber\\&&
-\ng \bigg(\frac{3341}{54}-\frac{604}{9}\z3\bigg)
\bigg)
-a_3^3a_1 \ng \bigg(\frac{7843}{972}+\frac{2915}{81}\z3
-\ng \bigg(
\frac{32791}{1458}
-\frac{5204}{243}\z3\bigg)
\bigg)
\nonumber\\&&
-a_3^2 (\yd^2+\yu^2) (427-96 \z3)
-a_3^2 \yd \yu \bigg(\frac{4282}{9}-\frac{400}{3}\z3\bigg)
-\frac{133}{3}a_3^2 (\yu +\yd )\yl 
\nonumber\\&&
+a_3^2a_1  \yu \bigg(\frac{1283}{36}+\frac{8}{3}\z3\bigg)
+a_3^2a_1  \yd \bigg(\frac{1487}{36}+\frac{104}{3}\z3\bigg)
+a_3^2a_2  (\yu+\yd)\bigg(\frac{473}{4}+72 \z3\bigg)
\nonumber\\&&
-a_3^2a_1^2  \ng\bigg(\frac{4711}{1458}
+\frac{275}{486}\z3
-\ng \bigg(
\frac{46783}{2187}
-\frac{5500}{729}\z3\bigg)\bigg)
-a_3^2a_2^2 \ng \bigg(
\frac{3764}{27}
-\frac{1075}{18}\z3
\nonumber\\&&
-\ng \bigg(
\frac{937}{27}
-\frac{100}{9}\z3\bigg)
\bigg)
-\frac{23}{12}a_3^2a_1 a_2  \ng 
-36 a_3 \lambda^2 (\yd+\yu)
+30 a_3 \lambda (\yd^2+\yu^2)
\nonumber\\&&
+a_3 (\yu^3+\yd^3) \bigg(\frac{423}{4}+6 \z3\bigg)
+a_3 \yu \yd (\yu+\yd) \bigg(\frac{1171}{12}+8 \z3\bigg)
+\frac{181}{8}a_3 \yl \bigg(\yu^2 +\yd^2 \bigg)
\nonumber\\&&
+a_3 \yd \yl \yu \bigg(\frac{515}{36}+\frac{16}{3}\z3\bigg)
+\frac{135}{8}a_3 \yl^2 (\yu +\yd)
-a_3a_2  \yl( \yu+\yd ) \bigg(9 \z3+\frac{295}{16}\bigg)
\nonumber\\&&
-a_3a_2  \yd \yu \bigg(\frac{1895}{16}+3 \z3\bigg)
-a_3a_1  \yl \yu \bigg(\frac{3649}{144}-9 \z3\bigg)
-a_3a_1  \yd \yl \bigg(\frac{4201}{144}-9 \z3\bigg)
\nonumber\\&&
-a_3a_1  \yu^2 \bigg(\frac{3641}{96}+\frac{7}{2}\z3\bigg)
-a_3a_1  \yd \yu \bigg(\frac{19033}{432}-\frac{25}{9}\z3\bigg)
-a_3a_1  \yd^2 \bigg(\frac{2869}{96}+\frac{27}{2}\z3\bigg)
\nonumber\\&&
-a_3a_2  (\yu^2+\yd^2) \bigg(\frac{3201}{32}+\frac{45}{2}\z3\bigg)
-\frac{21}{8}a_3a_2^2   \yl \ng
-\frac{253}{72}a_3a_1^2  \yl  \ng
\nonumber\\&&
-a_3a_1^2  \yu \bigg(\frac{26327}{5184}-\frac{19}{36}\z3+\frac{41995}{1944}\ng\bigg)
+a_3a_1^2  \yd \bigg(\frac{1657}{5184}+\frac{7}{36}\z3-\frac{26563}{1944}\ng\bigg)
\nonumber\\&&
+a_3a_1 a_2  \yd \bigg(\frac{775}{96}+\frac{15}{2}\z3\bigg)
-a_3a_1 a_2  \yu \bigg(\frac{77}{96}-\frac{45}{2}\z3\bigg)
+a_3a_2^2  (\yu+\yd) \bigg(\frac{31007}{192}
\nonumber\\&&
-\frac{117}{4}\z3
-\frac{755}{24}\ng\bigg)
+a_3a_1 a_2^2  \ng \bigg(\frac{9613}{5184}+\frac{5}{108}\z3+\ng\bigg(\frac{583}{324}-\frac{58}{27}\z3\bigg)\bigg)
\nonumber\\&&
+a_3a_1^2 a_2 \ng \bigg(\frac{13559}{5184}
-\frac{101}{108}\z3+\ng \bigg(\frac{2087}{972}-\frac{218}{81}\z3\bigg)\bigg)
+a_3a_1^3  \ng\bigg(
\frac{27491}{15552}
-\frac{1165}{2916}\z3
\nonumber\\&&
+\ng \bigg(
\frac{195455}{26244}
-\frac{21550}{2187}\z3\bigg)
+\frac{42350}{19683}\ng^2
\bigg)
-a_3a_2^3 \ng \bigg(\frac{283073}{5184}+\frac{385}{12}\z3
\nonumber\\&&
-\ng \bigg(\frac{7733}{324}+\frac{110}{3}\z3\bigg)
-\frac{154}{81}\ng^2
\bigg)\,,\label{eq:beta3}\\
\beta_2^{(4)}&&=
a_2^4 \bigg(
\frac{261826631}{62208}
+\frac{56045}{36}\z3
-\ng \bigg(\frac{9575323}{2592}
+\frac{4199}{18}\z3\bigg)
+\ng^2 \bigg(
\frac{28037}{54}
+148 \z3
\bigg)
\nonumber\\&&
+ \frac{1772}{243}\ng^3
\bigg)
+a_2^3 \bigg(\yu+\yd+\frac{\yl}{3}\bigg) \bigg(
\frac{938377}{2304}
-\frac{251}{6}\z3
-\ng \bigg(\frac{9119}{144}
-\frac{2}{3}\z3\bigg)
\bigg)
\nonumber\\&&
-a_1 a_2^3 \bigg(\frac{555089}{6912}
+\frac{169}{36}\z3
-\ng \bigg(
\frac{49589}{2592}
-\frac{767}{54}\z3\bigg)
-\ng^2 \bigg(
\frac{1391}{162}
-\frac{244}{27}\z3\bigg)
\bigg)
\nonumber\\&&
+a_2^3 a_3\ng \bigg(
\frac{6113}{216}
-\frac{1180}{9}\z3
+\ng \bigg(
\frac{2782}{27}
-\frac{976}{9}\z3\bigg)
\bigg)
- a_2^3 \lambda \bigg(
\frac{2857}{48}
+\frac{1}{3}\ng\bigg)
\nonumber\\&&
+a_1^2 a_2^2 \bigg(
\frac{1205}{6912}
-\frac{65}{36}\z3
+\ng \bigg(
\frac{55769}{2592}
-\frac{257}{54}\z3\bigg)
+\ng^2 \bigg(
\frac{2659}{486}
-\frac{20}{81}\z3\bigg)
\bigg)
\nonumber\\&&
-a_2^2 a_3^2 \ng\bigg(
\frac{3548}{9}
-\frac{440}{3}\z3
-\ng \bigg(
\frac{961}{27}
+\frac{200}{9}\z3\bigg)
\bigg)
+a_2^2 a_3 (\yu+\yd) \bigg(
\frac{361}{3}
+14 \z3
\bigg)
\nonumber\\&&
-a_2^2 \yd \yu \bigg(\frac{16735}{64}
+\frac{57}{2}\z3\bigg)
-a_2^2 \yl^2 \bigg(\frac{54931}{1152}
-\frac{59}{12}\z3\bigg)
-a_2^2 (\yu^2+\yd^2) \bigg(\frac{30213}{128}
-\frac{63}{4}\z3\bigg)
\nonumber\\&&
+a_1 a_2^2 \yu \bigg(
\frac{102497}{1152}
+\frac{35}{3}\z3\bigg)
-a_2^2 \yl \yu \bigg(
\frac{7007}{96}
+5 \z3
\bigg)
-a_2^2 \yd \yl \bigg(
\frac{8927}{96}
-\z3
\bigg)
\nonumber\\&&
+a_1 a_2^2 \yl \bigg(
\frac{15341}{384}
+8 \z3
\bigg)
+a_1 a_2^2 \yd \bigg(
\frac{90029}{1152}
+\frac{5}{3}\z3\bigg)
+\frac{75}{2} a_2^2 \lambda \bigg(\yu+\yd+\frac{\yl}{3}\bigg)
+\frac{363}{4} a_2^2 \lambda^2
\nonumber\\&&
-\frac{115}{8} a_1 a_2^2 \lambda
-\frac{161}{36} a_1 a_2^2 a_3 \ng
-52 a_2 \lambda^3
-75 a_2 \lambda^2 \bigg(\yu+\yd+\frac{\yl}{3}\bigg)
+\frac{1}{6} a_1 a_2 \lambda \yl
\nonumber\\&&
-a_1 a_2 a_3 \yd \bigg(
\frac{5}{3}
-26 \z3
\bigg)
+\frac{9}{2} a_1 a_2 \lambda \yu
+\frac{13}{2} a_2 \lambda \yl^2
+\frac{17}{2} a_1 a_2 \lambda \yd
+\frac{39}{2} a_2 \lambda (\yu^2+\yd^2)
\nonumber\\&&
+\frac{41}{2} a_2 \yl^2 (\yu+\yd)
+\frac{91}{4} a_2 \yl (\yu^2+\yd^2)
+\frac{109}{4} a_1 a_2 \lambda^2
-a_2 a_3 \yl(\yu+\yd) \bigg(
\frac{33}{2}
-16 \z3
\bigg)
\nonumber\\&&
+a_2 \yu \yd \yl \bigg(
\frac{113}{6}
-4 \z3\bigg)
-a_2 a_3 (\yu^2+\yd^2) \bigg(
\frac{239}{4}
-36 \z3
\bigg)
- a_1^2 a_2 \lambda \bigg(
\frac{377}{144}
+\frac{5}{27}\ng
\bigg)
\nonumber\\&&
-a_2 a_3 \yd \yu \bigg(
\frac{739}{6}
-152 \z3
\bigg)
+a_2 \yd \yu (\yu+\yd) \bigg(
\frac{3265}{32}
-6 \z3
\bigg)
-a_1 a_2 \yl \yu \bigg(
\frac{5401}{288}
-12 \z3
\bigg)
\nonumber\\&&
+a_2 \yl^3 \bigg(
\frac{269}{32}
+\frac{3}{2}\z3\bigg)
-a_1 a_2 \yd \yl \bigg(\frac{5471}{288}
+\frac{7}{3}\z3\bigg)
+a_2 (\yu^3+\yd^3) \bigg(
\frac{2143}{32}
+\frac{9}{2}\z3\bigg)
\nonumber\\&&
-a_1^2 a_2 \yu \bigg(\frac{40849}{6912}
-\frac{23}{6}\z3
+\ng \bigg(\frac{26515}{1296}
-\frac{10}{3}\z3\bigg)
\bigg)
-a_1 a_2 \yd^2 \bigg(\frac{15937}{384}
+\frac{33}{4}\z3\bigg)
\nonumber\\&&
+a_2 a_3^2 (\yu+\yd) \bigg(
\frac{823}{6}
-52 \z3
-\ng \bigg(\frac{86}{3}
+\frac{32}{3}\z3\bigg)\bigg)
-a_1 a_2 \yd \yu \bigg(\frac{33959}{576}
+\frac{40}{3}\z3\bigg)
\nonumber\\&&
+a_1^2 a_2 \yd \bigg(
\frac{10607}{6912}
+\frac{1}{6}\z3
-\ng \bigg(\frac{17447}{1296}
-\frac{70}{27}\z3\bigg)\bigg)
-a_1 a_2 \yl^2 \bigg(\frac{10309}{1152}
+\frac{49}{12}\z3\bigg)
\nonumber\\&&
-a_1 a_2 \yu^2 \bigg(\frac{15805}{384}
+\frac{51}{4}\z3\bigg)
-a_1^2 a_2 \yl \bigg(\frac{75739}{6912}
-\frac{77}{18}\z3
+\ng \bigg(\frac{14149}{1296}
-\frac{10}{27}\z3\bigg)
\bigg)
\nonumber\\&&
-a_1 a_2 a_3 \yu \bigg(\frac{199}{9}
-\frac{94}{3}\z3\bigg)
-a_1 a_2 a_3^2 \ng\bigg(\frac{16}{81}
-\frac{88}{27}\z3
-\ng \bigg(
\frac{2201}{243}
-\frac{824}{81}\z3\bigg)\bigg)
\nonumber\\&&
+a_1^2 a_2 a_3 \ng\bigg(
\frac{18905}{1944}
-\frac{796}{81}\z3
+\ng \bigg(
\frac{4174}{729}
-\frac{1744}{243}\z3\bigg)
\bigg)
-a_2 a_3^3\ng \bigg(
\frac{10487}{27}
+\frac{1760}{9}\z3
\nonumber\\&&
-\ng \bigg(
\frac{2828}{27}
+\frac{3520}{27}\z3\bigg)
- \frac{1232}{243}\ng^2
\bigg)
+a_1^3 a_2 \bigg(
\frac{61357}{62208}
-\frac{7}{36}\z3
+\ng \bigg(
\frac{208129}{23328}
-\frac{3155}{486}\z3\bigg)
\nonumber\\&&
+\ng^2 \bigg(
\frac{7085}{1458}
-\frac{3700}{729}\z3\bigg)
+ \frac{7700}{6561}\ng^3
\bigg)\,,\label{eq:beta2}\\
\beta_1^{(4)}&&=
a_1^4 \bigg(
\frac{64381}{62208}
-\frac{19}{36}\z3
+\ng \bigg(
\frac{1026515}{23328}
-\frac{60805}{4374}\z3\bigg)
-\ng^2 \bigg(\frac{927475}{39366}
-\frac{440500}{6561}\z3\bigg)
\nonumber\\&&
+ \frac{731500}{59049}\ng^3
\bigg)
+a_1^3 a_2 \bigg(
\frac{7123}{2304}
-\frac{7}{12}\z3
+\ng \bigg(
\frac{176093}{7776}
-\frac{833}{162}\z3\bigg)
+\ng^2 \bigg(
\frac{18335}{1458}
-\frac{3700}{243}\z3\bigg)\bigg)
\nonumber\\&&
+a_1^3 a_3\ng \bigg(
\frac{584713}{17496}
-\frac{7676}{729}\z3
+\ng \bigg(
\frac{380570}{6561}
-\frac{172400}{2187}\z3\bigg)\bigg)
- a_1^3 \lambda \bigg(
\frac{323}{48}
+\frac{5}{9}\ng
\bigg)
\nonumber\\&&
-a_1^3 \yd \bigg(
\frac{11}{54}\z3
-\frac{74651}{62208}
+\ng \bigg(\frac{246955}{11664}
-\frac{130}{81}\z3\bigg)\bigg)
-a_1^3 \yl \bigg(\frac{51127}{6912}
+\frac{529}{18}\z3
\nonumber\\&&
+\ng \bigg(\frac{64465}{1296}
+\frac{230}{27}\z3\bigg)
\bigg)
-a_1^3 \yu \bigg(\frac{250657}{62208}
+\frac{239}{54}\z3
+\ng \bigg(\frac{545515}{11664}
+\frac{530}{81}\z3\bigg)\bigg)
\nonumber\\&&
-a_1^2 a_2^2 \bigg(\frac{46315}{2304}
+\frac{131}{12}\z3
-\ng \bigg(
\frac{77605}{7776}
-\frac{235}{162}\z3\bigg)
+\ng^2 \bigg(\frac{2021}{486}
-\frac{700}{81}\z3\bigg)\bigg)
\nonumber\\&&
-a_1^2 a_3^2 \bigg(\ng \bigg(\frac{2192}{729}
-\frac{12056}{243}\z3\bigg)
+\ng^2 \bigg(\frac{81197}{2187}
-\frac{56600}{729}\z3\bigg)\bigg)
-a_1^2 \yd^2 \bigg(\frac{47975}{1152}
+\frac{5}{4}\z3\bigg)
\nonumber\\&&
+a_1^2 a_2 \yd \bigg(
\frac{4145}{128}
+\frac{5}{6}\z3\bigg)
+a_1^2 a_3 \yd \bigg(
\frac{395}{27}
+10 \z3
\bigg)
+a_1^2 a_3 \yu \bigg(
\frac{503}{27}
-\frac{34}{3}\z3\bigg)
\nonumber\\&&
+a_1^2 a_2 \yu \bigg(
\frac{42841}{1152}
+\frac{187}{6}\z3\bigg)
-a_1^2 \yd \yl \bigg(\frac{11837}{864}
+\frac{25}{9}\z3\bigg)
-a_1^2 \yu^2 \bigg(\frac{145295}{1152}
-\frac{119}{12}\z3\bigg)
\nonumber\\&&
-a_1^2 \yd \yu \bigg(\frac{306401}{5184}
-\frac{125}{27}\z3\bigg)
-a_1^2 a_2 \yl \bigg(\frac{2901}{128}
-\frac{135}{2}\z3\bigg)
-a_1^2 \yl^2 \bigg(\frac{89161}{1152}
-\frac{155}{12}\z3\bigg)
\nonumber\\&&
-a_1^2 \yl \yu \bigg(\frac{70529}{864}
-\frac{1370}{9}\z3\bigg)
+\frac{23}{36} a_1^2 a_2 a_3 \ng
+\frac{59}{2} a_1^2 \lambda \yd
+\frac{73}{2} a_1^2 \lambda \yl
+\frac{107}{2} a_1^2 \lambda \yu
\nonumber\\&&
+\frac{141}{4} a_1^2 \lambda^2
-\frac{213}{8} a_1^2 a_2 \lambda
-52 a_1 \lambda^3
-63 a_1 \lambda^2 \yd
-61 a_1 \lambda^2 \yl
+80 a_1 \yd^2 \yl
-99 a_1 \lambda^2 \yu
\nonumber\\&&
+23a_1 a_2 a_3 \yd \bigg(
1+2 \z3\bigg)
+\frac{1}{2} a_1 a_2 \lambda \yl
+\frac{19}{2} a_1 \lambda \yd^2
+\frac{27}{2} a_1 a_2 \lambda \yu
+\frac{51}{2} a_1 a_2 \lambda \yd
+\frac{73}{2} a_1 \lambda \yl^2
\nonumber\\&&
+\frac{79}{2} a_1 \lambda \yu^2
+\frac{249}{4} a_1 \yd \yl^2
+\frac{327}{4} a_1 a_2 \lambda^2
-a_1 a_2 a_3 \yu \bigg(
\frac{367}{3}
-158 \z3
\bigg)
-a_1 a_3 \yd^2 \bigg(
\frac{497}{12}
-4 \z3
\bigg)
\nonumber\\&&
+\frac{633}{8} a_1 \yl^2 \yu
-a_1 a_2^2 \lambda \bigg(
\frac{841}{48}
+\frac{1}{3} \ng
\bigg)
-a_1 a_3 \yu^2 \bigg(
\frac{1429}{12}
-100 \z3
\bigg)
+\frac{2371}{24} a_1 \yl \yu^2
\nonumber\\&&
-a_1 a_2 \yd \yl \bigg(
\frac{2869}{32}
+47 \z3
\bigg)
-a_1 a_3 \yl \yu \bigg(
\frac{3061}{18}
-112 \z3
\bigg)
-a_1 a_3 \yd \yl \bigg(
\frac{3613}{18}
-112 \z3
\bigg)
\nonumber\\&&
-a_1 a_2 \yl \yu \bigg(
\frac{9713}{96}
+19 \z3
\bigg)
+a_1 \yd^3 \bigg(
\frac{1595}{32}
+\frac{5}{2}\z3\bigg)
+a_1 \yd \yu^2 \bigg(
\frac{29281}{288}
-\frac{14}{3}\z3\bigg)
\nonumber\\&&
+a_1 \yd^2 \yu \bigg(
\frac{32269}{288}
-\frac{14}{3}\z3\bigg)
+a_1 \yl^3 \bigg(
\frac{4123}{96}
+\frac{15}{2}\z3\bigg)
+a_1 \yu^3 \bigg(
\frac{4551}{32}
+\frac{17}{2}\z3\bigg)
\nonumber\\&&
+a_1 \yd \yl \yu \bigg(
\frac{5933}{216}
-\frac{28}{9}\z3\bigg)
-a_1 a_2 \yd^2 \bigg(\frac{14977}{128}
+\frac{33}{4}\z3\bigg)
-a_1 a_2 \yl^2 \bigg(\frac{31141}{384}
+\frac{55}{4}\z3\bigg)
\nonumber\\&&
+a_1 a_2^2 \yd \bigg(
\frac{128359}{768}
-\frac{92}{3}\z3
-\ng \bigg(\frac{1001}{48}
-\frac{10}{3}\z3\bigg)
\bigg)
-a_1 a_2 \yd \yu \bigg(\frac{30655}{192}
+\frac{127}{2}\z3\bigg)
\nonumber\\&&
+a_1 a_3^2 \yd \bigg(
\frac{4211}{54}
+\frac{116}{3}\z3
-\ng \bigg(\frac{1898}{81}
+\frac{32}{3}\z3\bigg)
\bigg)
-a_1 a_2 \yu^2 \bigg(\frac{23821}{128}
+\frac{213}{4}\z3\bigg)
\nonumber\\&&
+a_1 a_2^2 \yl \bigg(
\frac{560095}{2304}
-\frac{163}{3}\z3
-\ng \bigg(\frac{5297}{144}
-\frac{14}{3}\z3\bigg)
\bigg)
-a_1 a_3 \yd \yu \bigg(\frac{6901}{54}
-\frac{1208}{9}\z3\bigg)
\nonumber\\&&
-a_1 a_2^3 \bigg(
\frac{5054111}{20736}
-\frac{101}{4}\z3
-\ng \bigg(
\frac{201985}{7776}
+\frac{83}{18}\z3\bigg)
-\ng^2 \bigg(
\frac{8027}{486}
+\frac{220}{9}\z3\bigg)
- \frac{308}{243}\ng^3
\bigg)
\nonumber\\&&
-a_1 a_3^3\ng \bigg(
\frac{115357}{243}
+\frac{19360}{81}\z3
-\ng \bigg(
\frac{29348}{243}
+\frac{42560}{243}\z3\bigg)
-\frac{13552}{2187}\ng^2
\bigg)
\nonumber\\&&
+a_1 a_2^2 \yu \bigg(
\frac{784625}{2304}
-\frac{232}{3}\z3
-\ng \bigg(\frac{7183}{144}
-\frac{26}{3}\z3\bigg)
\bigg)
+a_1 a_2^2 a_3\ng \bigg(
\frac{13987}{648}
-\frac{476}{27}\z3
\nonumber\\&&
+\ng \bigg(
\frac{1166}{81}
-\frac{464}{27}\z3\bigg)
\bigg)
+a_1 a_3^2 \yu \bigg(
\frac{13799}{54}
-\frac{700}{3}\z3
-\ng \bigg(\frac{4034}{81}
+\frac{32}{3}\z3\bigg)
\bigg)
\nonumber\\&&
-a_1 a_2 a_3^2\ng \bigg(
\frac{16}{27}
-\frac{88}{9}\z3
-\ng \bigg(
\frac{2201}{81}
-\frac{824}{27}\z3\bigg)\bigg)\,.
\label{eq:beta1}
\end{eqnarray}
We find full agreement with the results of Refs.~\cite{Davies:2019onf,Bednyakov:2021qxa}, computed with the help of WCCs.

As a by-product of our calculation, we obtain the four-loop renormalization constants of all SM gauge fields in general gauge, which are completely new results.
The full results may be found in the ancillary file.
Here, we present the anomalous dimensions of the gauge fields in Landau gauge:
\begin{eqnarray}
\gamma_{GG}^{(4)}&&=
a_3^4 \bigg(
-\frac{10596127}{768}
+\frac{1012023}{256}\z3
-\frac{8019}{32}\z4 
- \frac{40905}{4}\z5
+\ng \bigg(\frac{23350603}{2592}-\frac{387649}{108}\z3
\nonumber\\&&
+\frac{8955}{8}\z4 +3355 \z5\bigg)
-\ng^2 \bigg(\frac{86066}{81}+\frac{8068}{81}\z3+132 \z4\bigg)
-\ng^3 \bigg(\frac{17708}{729}-\frac{64}{3}\z3\bigg)
\bigg)
\nonumber\\&&
-a_3^3 (\yu+\yd) \bigg(\frac{67117}{72}-434 \z3-\ng \bigg(\frac{4159}{27}-16 \z3\bigg)\bigg)
+a_3^3 a_1 \ng\bigg(
\bigg(\frac{1263911}{7776}
\nonumber\\&&
+\frac{33055}{162}\z3
-121 \z4-\frac{550}{3}\z5 \bigg)
-\ng \bigg(\frac{133595}{2916}-\frac{2828}{243}\z3-\frac{44}{3}\z4\bigg)
\bigg)
\nonumber\\&&
+a_3^3 a_2 \ng\bigg(
\bigg(\frac{114901}{288}+\frac{3005}{6}\z3
-297 \z4-450 \z5\bigg)
-\ng \bigg(\frac{12865}{108}-\frac{388}{9}\z3-36 \z4\bigg)
\bigg)
\nonumber\\&&
+a_3^2 (\yu^2+\yd^2) \bigg(\frac{3817}{16}-\frac{111}{2}\z3\!\bigg)
+a_3^2 \yd \yu \bigg(\frac{16067}{72}-\frac{283}{3}\z3\!\bigg)
-a_3^2 (\yu+\yd) \yl \bigg(\frac{8}{3}-12 \z3\!\bigg)
\nonumber\\&&
+a_3^2 a_1 \yu \bigg(\frac{3277}{144}-\frac{271}{6}\z3\bigg)
+a_3^2 a_1 \yd \bigg(\frac{985}{144}-\frac{283}{6}\z3\bigg)
+a_3^2 a_2 (\yu+\yd) \bigg(\frac{583}{16}-\frac{279}{2}\z3\bigg)
\nonumber\\&&
+a_3^2 a_1^2\ng  \bigg(
\bigg(\frac{737543}{46656}
+\frac{36623}{1944}\z3+\frac{11}{12}\z4-\frac{685}{18}\z5\!\bigg)
+\ng \bigg(\frac{104929}{4374}-\frac{27170}{729}\z3+\frac{110}{9}\z4\! \bigg)\!
\bigg)
\nonumber\\&&
-a_3^2 a_2^2\ng \bigg(
\bigg(\frac{425131}{1728}-\frac{33797}{72}\z3+\frac{387}{4}\z4 +\frac{225}{2}\z5\bigg)
-\ng \bigg(\frac{1735}{54}-\frac{494}{9}\z3+18 \z4\bigg)
\bigg)
\nonumber\\&&
+a_3^2 a_1 a_2 \ng \bigg(\frac{455}{96}+\frac{37}{4}\z3-15 \z5\bigg)
+36 a_3  (\yu+\yd)\lambda^2
-30 a_3  (\yu^2+\yd^2)\lambda
+\frac{21}{8}a_3 a_2^2 \yl \ng 
\nonumber\\&&
-\frac{181}{8}a_3 (\yu^2+\yd^2) \yl 
-\frac{135}{8}a_3 (\yu+\yd) \yl^2 
+\frac{253}{72}a_3 a_1^2  \yl \ng
-a_3 (\yu^3+\yd^3) \bigg(\frac{423}{4}+6 \z3\bigg)
\nonumber\\&&
-a_3 (\yu+\yd) \yd \yu \bigg(\frac{1171}{12}+8 \z3\bigg)
-a_3 \yd \yl \yu \bigg(\frac{515}{36}+\frac{16}{3}\z3\bigg)
+a_3 a_1 \yu^2 \bigg(\frac{3641}{96}+\frac{7}{2}\z3\bigg)
\nonumber\\&&
+a_3 a_1 \yu \yl \bigg(\frac{3649}{144}-9 \z3\!\bigg)
+a_3 a_1 \yd \yl \bigg(\frac{4201}{144}-9 \z3\!\bigg)
+a_3 a_1 \yu\yd  \bigg(\frac{19033}{432}-\frac{25}{9}\z3\!\bigg)
\nonumber\\&&
+a_3 a_1 \yd^2 \bigg(\frac{2869}{96}+\frac{27}{2}\z3\!\bigg)
+a_3 a_2 (\yu+\yd) \yl \bigg(\frac{295}{16}+9 \z3\bigg)
+a_3 a_2 \yd \yu \bigg(\frac{1895}{16}+3 \z3\bigg)
\nonumber\\&&
-a_3 a_1^2 \yd \bigg(\frac{1657}{5184}-\ng \frac{26563}{1944}+\frac{7}{36}\z3\bigg)
+a_3 a_1^2 \yu \bigg(\frac{26327}{5184}+\ng \frac{41995}{1944}-\frac{19}{36}\z3\bigg)
\nonumber\\&&
+a_3 a_2 (\yu^2+\yd^2) \bigg(\frac{3201}{32}
+\frac{45}{2}\z3\bigg)
-a_3 a_1 a_2^2 \bigg(\ng \bigg(\frac{9613}{5184}+\frac{5}{108}\z3\bigg)+\ng^2 \bigg(\frac{583}{324}-\frac{58}{27}\z3\bigg)\bigg)
\nonumber\\&&
-a_3 a_1 a_2 \yd \bigg(\frac{775}{96}+\frac{15}{2}\z3\bigg)
+a_3 a_1 a_2 \yu \bigg(\frac{77}{96}-\frac{45}{2}\z3\bigg)
-a_3 a_1^2 a_2 \ng\bigg(\bigg(\frac{13559}{5184}-\frac{101}{108}\z3\bigg)
\nonumber\\&&
+\ng \bigg(\frac{2087}{972}-\frac{218}{81}\z3\bigg)\bigg)
-a_3 a_2^2 (\yu+\yd) \bigg(\frac{31007}{192}-\frac{117}{4}\z3-\frac{755}{24}\ng\bigg)
\nonumber\\&&
-a_3 a_1^3\ng \bigg(
\frac{27491}{15552}-\frac{1165}{2916}\z3
+\ng \bigg(\frac{195455}{26244}-\frac{21550}{2187}\z3\bigg)
+\frac{42350}{19683}\ng^2
\bigg)
\nonumber\\&&
+a_3 a_2^3 \ng\bigg(
\frac{283073}{5184}+\frac{385}{12}\z3
+\ng \bigg(-\frac{7733}{324}-\frac{110}{3}\z3\bigg)
- \frac{154}{81}\ng^2
\bigg)
,\\
\gamma_{WW}^{(4)}&&=
- a_2^4 \bigg(
\frac{140654297}{62208}
-\frac{1793}{6}\z3
-\frac{129}{8}\z4
+ \frac{79405}{32}\z5
-\ng \bigg(
\frac{6494521}{2592}
-\frac{10073}{9}\z3
\nonumber\\&&
+384 \z4
+1070 \z5
\bigg)
+\ng^2 \bigg(
\frac{24125}{54}
+\frac{68}{3}\z3
+72 \z4
\bigg)
+\ng^3 \bigg(
\frac{3764}{243}
-\frac{128}{9}\z3
\bigg)
\bigg)
\nonumber\\&&
+a_2^3 a_3\ng \bigg(
\bigg(
\frac{85231}{216}
+\frac{4648}{9}\z3
-344 \z4
-400 \z5
\bigg)
-\ng \bigg(
\frac{5530}{27}
-\frac{592}{9}\z3
-64 \z4
\bigg)
\bigg)
\nonumber\\&&
+a_1 a_2^3 \bigg(
\frac{554021}{6912}
-\frac{35}{9}\z3
- \frac{43}{4}\z4
- \frac{25}{2}\z5
+\ng \bigg(
\frac{17419}{2592}
+\frac{2537}{54}\z3
- \frac{80}{3}\z4
- \frac{100}{3}\z5
\bigg)
\nonumber\\&&
-\ng^2 \bigg(
\frac{2765}{162}
-\frac{148}{27}\z3
- \frac{16}{3}\z4
\bigg)
\bigg)
+ a_2^3 \lambda \bigg(
\frac{2425}{48}
+\frac{\ng}{3}
\bigg)
-a_2^3 \bigg(\yu+\yd+\frac{\yl}{3}\bigg) \bigg(
\frac{818785}{2304}
\nonumber\\&&
-\frac{523}{3}\z3
-\ng \bigg(\frac{14519}{144}
-\frac{50}{3}\z3\bigg)
\bigg)
- a_1^2 a_2^2 \bigg(
\frac{14995}{6912}
-\frac{439}{72}\z3
-\frac{1}{8}\z4
+ \frac{15}{4}\z5
+\ng \bigg(
\frac{1297}{864}
\nonumber\\&&
-\frac{14}{9}\z3
-2 \z4
+ \frac{70}{9}\z5
\bigg)
-\ng^2 \bigg(
\frac{1787}{162}
-\frac{1300}{81}\z3
+ \frac{40}{9}\z4\!
\bigg)
\bigg)
-a_2^2 a_3^2 \ng\bigg(
\frac{5852}{9}
-1360 \z3
\nonumber\\&&
+264 \z4
+400 \z5
-\ng \bigg(
\frac{749}{9}
-\frac{1256}{9}\z3
+32 \z4
\bigg)\bigg)
+a_1 a_2^2 a_3 \ng \bigg(
\frac{1025}{108}
+\frac{148}{9}\z3
-\frac{80}{3}\z5
\bigg)
\nonumber\\&&
-a_1 a_2^2 \yu \bigg(
\frac{2525}{1152}
+\frac{235}{6}\z3\bigg)
-a_1 a_2^2 \yd \bigg(
\frac{22361}{1152}
+\frac{91}{6}\z3\bigg)
+a_2^2 a_3 (\yu +\yd )\bigg(
\frac{221}{3}
-134 \z3
\bigg)
\nonumber\\&&
-\frac{225}{4} a_2^2 \lambda^2
+a_2^2 \yu \yl \bigg(
\frac{671}{96}
+11 \z3
\bigg)
+a_2^2 \yu \yd \bigg(\frac{4863}{64}
+\frac{63}{2}\z3\bigg)
+a_2^2 \yd \yl \bigg(
\frac{2591}{96}
+5 \z3
\bigg)
\nonumber\\&&
+a_2^2 (\yu^2+\yd^2) \bigg(\frac{10101}{128}
+\frac{27}{4}\z3\bigg)
+a_2^2 \yl^2 \bigg(\frac{19939}{1152}
+\frac{7}{12}\z3\bigg)
+a_1 a_2^2 \yl \bigg(\frac{1165}{128}
-\frac{57}{2}\z3\bigg)
\nonumber\\&&
-\frac{75}{2} a_2^2 \lambda \bigg(\yu +\yd+\frac{\yl}{3} \bigg)
+\frac{91}{8} a_1 a_2^2 \lambda
-a_1^3 a_2 \bigg(
\frac{61357}{62208}
-\frac{7}{36}\z3
+\ng \bigg(
\frac{208129}{23328}
-\frac{3155}{486}\z3\bigg)
\nonumber\\&&
+\ng^2 \bigg(
\frac{7085}{1458}
-\frac{3700}{729}\z3\bigg)
+\frac{7700}{6561}\ng^3 
\bigg)
+a_2 a_3^3 \ng\bigg(
\frac{10487}{27}
+\frac{1760}{9}\z3
-\ng \bigg(
\frac{2828}{27}
+\frac{3520}{27}\z3\bigg)
\nonumber\\&&
- \frac{1232}{243}\ng^2
\bigg)
+a_1 a_2 a_3^2\ng \bigg(
\frac{16}{81}
-\frac{88}{27}\z3
-\ng \bigg(
\frac{2201}{243}
-\frac{824}{81}\z3\bigg)
\bigg)
-a_1^2 a_2 a_3 \ng\bigg(
\frac{18905}{1944}
\nonumber\\&&
-\frac{796}{81}\z3
+\ng \bigg(
\frac{4174}{729}
-\frac{1744}{243}\z3\bigg)
\bigg)
-a_1^2 a_2 \yd \bigg(
\frac{10607}{6912}
+\frac{\z3}{6}
-\ng \bigg(\frac{17447}{1296}
-\frac{70}{27}\z3\bigg)\bigg)
\nonumber\\&&
+a_1^2 a_2 \yl \bigg(\frac{75739}{6912}
-\frac{77}{18}\z3
+\ng \bigg(\frac{14149}{1296}
-\frac{10}{27}\z3\bigg)
\bigg)
+a_1 a_2 \yu^2 \bigg(\frac{15805}{384}
+\frac{51}{4}\z3\bigg)
\nonumber\\&&
+a_1^2 a_2 \yu \bigg(\frac{40849}{6912}
-\frac{23}{6}\z3
+\ng \bigg(\frac{26515}{1296}
-\frac{10}{3}\z3\bigg)
\bigg)
+a_1 a_2 \yd \yu \bigg(\frac{33959}{576}
+\frac{40}{3}\z3\bigg)
\nonumber\\&&
+a_2 a_3^2 (\yu+\yd) \bigg(
52 \z3
-\frac{823}{6}
+\ng \bigg(\frac{86}{3}
+\frac{32}{3}\z3\bigg)\bigg)
+a_1 a_2 \yd^2 \bigg(\frac{15937}{384}
+\frac{33}{4}\z3\bigg)
\nonumber\\&&
+a_1 a_2 \yl^2 \bigg(\frac{10309}{1152}
+\frac{49}{12}\z3\bigg)
+a_1 a_2 a_3 \yu \bigg(\frac{199}{9}
-\frac{94}{3}\z3\bigg)
+a_2 a_3 \yl(\yu +\yd ) \bigg(
\frac{33}{2}
-16 \z3
\bigg)
\nonumber\\&&
+a_2 a_3 \yd \yu \bigg(
\frac{739}{6}
-152 \z3
\bigg)
-a_2 \yd \yu (\yu +\yd )\bigg(
\frac{3265}{32}
-6 \z3
\bigg)
+a_1 a_2 \yl \yu \bigg(
\frac{5401}{288}
-12 \z3
\bigg)
\nonumber\\&&
-a_2 (\yu^3+ \yd^3) \bigg(
\frac{2143}{32}
+\frac{9}{2}\z3\bigg)
+a_2 a_3 (\yu^2+ \yd^2) \bigg(
\frac{239}{4}
-36 \z3
\bigg)
+a_1 a_2 \yd \yl \bigg(\frac{5471}{288}
+\frac{7}{3}\z3\bigg)
\nonumber\\&&
-a_2 \yl^3 \bigg(
\frac{269}{32}
+\frac{3}{2}\z3\bigg)
-a_2 \yd \yl \yu \bigg(
\frac{113}{6}
-4 \z3
\bigg)
+a_1 a_2 a_3 \yd \bigg(
\frac{5}{3}
-26 \z3
\bigg)
-\frac{1}{6} a_1 a_2 \lambda \yl
\nonumber\\&&
-\frac{41}{2} a_2 \yl^2(\yu +\yd )
-\frac{91}{4} a_2 \yl (\yu^2+ \yd^2)
+75 a_2 \lambda^2 \bigg(\yu+\yd+\frac{\yl}{3}\bigg)
-\frac{9}{2} a_1 a_2 \lambda \yu
-\frac{13}{2} a_2 \lambda \yl^2
\nonumber\\&&
-\frac{17}{2} a_1 a_2 \lambda \yd
+a_1^2 a_2 \lambda \bigg(
\frac{377}{144}
+\frac{5}{27}\ng
\bigg)
-\frac{39}{2} a_2 \lambda (\yu^2+ \yd^2)
-\frac{109}{4} a_1 a_2 \lambda^2
+52 a_2 \lambda^3
.
\end{eqnarray}
The four-loop anomalous dimension $\gamma_{BB}^{(4)}$ of the $B$ field can be obtained from the expression for $\beta_1^{(4)}$  in Eq.~(\ref{eq:beta1}) with the help of Eq.~(\ref{eq:bftog}) because, in this case, we do not have any additional vertices in the background gauge.
The result is simply $\gamma_{BB}^{(4)}=-\beta_1^{(4)}$.

\section{Conclusion}
\label{sec:five}

We directly computed all gauge coupling beta functions and gauge field anomalous dimensions at four loops the SM.
The main challenge of the calculation was the treatment of $\gamma_5$.
We used the reading point method~\cite{Korner:1991sx}, choosing the internal vertices as the reading points, as suggested in Ref.~\cite{Bednyakov:2015ooa}.
The resulting expressions for the gauge coupling beta functions fully agree with Refs.~\cite{Davies:2019onf,Bednyakov:2021qxa}.
Our results for the gauge field anomalous dimensions are new and complete the four-loop renormalization of the SM gauge sector in the unbroken phase.

\subsection*{Acknowledgments}
This work was supported by the German Research Foundation DFG through Grant No.~
KN~365/16-1.


\end{document}